\documentclass[useAMS,usenatbib,fleqn]{mn2e}
\usepackage{amsmath,amssymb, aas_macros,times}
\usepackage{graphicx}
\usepackage[usenames,dvipsnames]{xcolor}
\usepackage{color}
\newcommand{\beq}{\begin{equation}}
\newcommand{\eeq}{\end{equation}}
\newcommand{\beqarray}{\begin{eqnarray}}
\newcommand{\eeqarray}{\end{eqnarray}}
\usepackage{float}

\begin{document}
\title[Photodisintegrated $\gamma$ and $\nu$ from GRB 130427A]{Photodisintegrated gamma rays and neutrinos from heavy nuclei in the gamma-ray burst jet of GRB 130427A} 

\author[Joshi et al.]{
Jagdish C.\ Joshi$^{1,2}$\thanks{jagdish@rri.res.in}, 
Soebur Razzaque$^{2}$\thanks{srazzaque@uj.ac.za} and 
Reetanjali Moharana$^{2}$\thanks{reetanjalim@uj.ac.za} \\
$^{1}$ Raman Research Institute, Sadashivanagar, Bangalore 560080, India\\
$^{2}$ Department of Physics, University of Johannesburg, PO Box 524, Auckland Park 2006, South Africa 
}

\bibliographystyle{mn2e}


\maketitle

\begin{abstract}
Detection of $\sim 0.1$-70~GeV prompt $\gamma$-ray emission from the exceptionally bright gamma-ray burst (GRB) 130427A by the {\it Fermi}-Large Area Telescope provides an opportunity to explore the physical processes of GeV $\gamma$-ray emission from the GRB jets. In this work we discuss interactions of Iron and Oxygen nuclei with observed keV-MeV photons in the jet of GRB 130427A in order to explain an additional, hard spectral component observed during 11.5-33 second after trigger. The photodisintegration time scale for Iron nuclei is comparable to or shorter than this duration. We find that $\gamma$ rays resulting from the Iron nuclei disintegration can account for the hard power-law component of the spectra in the $\sim 1$-70~GeV range, before the $\gamma\gamma \to e^\pm$ pair production with low-energy photons severely attenuates emission of higher energy photons. Electron antineutrinos from the secondary neutron decay, on the other hand, can be emitted with energies up to $\sim 2$~TeV. The flux of 
these neutrinos is low and consistent with non-detection of GRB~130427A by the IceCube Neutrino Observatory. The required total energy in the Iron nuclei for this hadronic model for GeV emission is $\lesssim 10$ times the observed total energy released in the prompt keV-MeV emission.
\end{abstract}
 
\date{\today}
\maketitle

\section{Introduction}

Gamma-ray burst (GRB) 130427A triggered the {\it Fermi}-Gamma-ray Burst Monitor (GBM) at time $T_0$ = 07:47:06:42 UTC~\citep{gbmtrigger} and it was followed-up by the {\it Fermi}-Large Area Telescope (LAT)~\citep{zhulat}. The GBM location of GRB 130427A was consistent with the {\it Swift}-BAT location~\citep{masellibat} and more than 50 telescopes subsequently observed this extraordinary burst. Located at a redshift of $z = 0.34$~\citep{130427z}, GRB~130427A is found to be associated with a type-Ic supernova SN~2013cq~\citep{grb_sn}, thus providing further evidence that core-collapse of massive stars are possibly the progenitors of long-duration GRBs. The isotropic-equivalent $\gamma$-ray energy of GRB~130427A is $E_{\gamma,\rm iso} = 8.1\times 10^{53}$~erg and the peak luminosity is $L_{\gamma,\rm iso} = 2.7\times 10^{53}$~erg/s, making it one of the most energetic GRBs ever detected~\citep{Maselli+2014}.

{\it Fermi}-LAT has detected GeV emission from GRB~130427A for almost a day~\citep{sci_grb_130427a} while the GBM 10-1000~keV emission lasted for $\sim 350$~s~\citep{Preece_grb130427A}. {\it Fermi}-LAT detections of a 73 GeV photon during the prompt phase at $T_0 + 19$~s and of a 95 GeV photon during the afterglow phase at $T_0 + 244$~s are particularly remarkable as they provide meaningful constraints on the GRB jet parameters and emission processes~\citep{sci_grb_130427a}. A minimum jet bulk Lorentz factor of $\Gamma_{\rm min} \sim 450$ is needed for GRB 130427A, assuming emission of the 73 GeV photon is unattenuated by the $\gamma\gamma\to e^\pm$ pair-production process in the internal-shocks scenario~\citep{sci_grb_130427a}. The 95 GeV photon, the most energetic detected yet from a GRB, provides strong constraints on the electron-synchrotron emission as the physical process for production of this energetic photon in the afterglow phase~\citep{sci_grb_130427a}.

While it is generally accepted now that temporally-extended GeV emission, long after the keV-MeV emission is over, is likely afterglow 
synchrotron and/or inverse Compton emission from a decelerating blastwave~\citep[see, e.g.,][]{Kumar+2009,Kumar+2010,Ghisellini+2010,Razzaque2010} as 
also discussed for GRB~130427A~\citep{Tam+2013,sci_grb_130427a,Beloborodov+2014}; it is far from clear what is the mechanism for prompt GeV emission. 
Both leptonic, primarily inverse Compton emission~\citep{Wang+2009,Bosnjak+2009,Peer+2012} and hadronic, proton-synchrotron~\citep{Razzaque+2010} and 
photohadronic interactions~\citep{Asano+2009}, have been considered in the literature \citep[see also, 
for reviews,][]{review_supp1}. 

In this letter we focus on photodisintegration of heavy nuclei
\begin{equation}
 A + \gamma \rightarrow A^* \rightarrow (A-1) + \gamma + n/p,
\label{photo}
\end{equation}
interacting with the observed keV-MeV photons in the prompt phase of GRB~130427A, to produce the observed GeV emission that is detected as an additional power-law component during the $T_0+11.5$~s to $T_0+33$~s interval~\citep{sci_grb_130427a}. Heavy nuclei such as the Iron (Fe) and Oxygen (O) can be entrapped in the GRB jet from the stellar envelope as the jet propagates inside the GRB progenitor star~\citep{Zhang+2003}. It was discussed by~\citet{wang} that these heavy nuclei can survive nuclear spallation due to nucleon-nucleon collisions and photodisintegration while interacting with the thermal photons in the GRB jet at the stellar Fe core radius of $\sim 10^9$~cm. In particular, the first $\sim 10$~s of the GRB jet could be rich in heavy nuclei~\citep{wang}. Non-thermal photons created in the internal-shocks, e.g., from synchrotron radiation of shock-accelerated electrons, outside the stellar envelope, however, can break down the shock-accelerated heavy nuclei~\citep{wang} and lead to high-energy $\
gamma$-ray emission~\citep{murase_pd}.

We discuss photodisintegration process using parameters of GRB~130427A in Sec.~\ref{sec:photo} and resulting $\gamma$-ray flux in Sec.~\ref{sec:gamma_flux}. We calculate neutrino flux from the photodisintegration process and its detectability in Sec.~\ref{sec:neutrino}, discuss our results and conclude in Sec.~\ref{sec:conclusion}.

\section{Photodisintegration in the GRB jet}
\label{sec:photo}

In this calculation we have used four frames of references: (i) The comoving GRB jet frame or wind rest frame denoted with superscript `$\prime$', (ii) the lab frame or GRB source frame denoted with superscript `$*$', (iii) the rest-frame of the nuclei denoted with superscript `$\prime\prime$', and (iv) the observer frame with no superscript. The energies in the observer frame, lab frame and comoving jet frame are related by the Lorentz boost factor $\Gamma$ of the bulk GRB outflow and redshift $z$ by the relation $E_\gamma = E_\gamma^*/(1+z) = \Gamma E_\gamma^\prime/(1+z)$. 

For calculation, we use fiducial parameter values $\Gamma = 10^3$, $R_{\rm in} = 2\times 10^{13}$~cm and $L_{\gamma,\rm iso} = 10^{52}$~erg/s for the 11.5-33.0~s interval of GRB~130427A, where $R_{\rm in}$ is the dissipation radius and $L_{\gamma,\rm iso}$ is the isotropic-equivalent $\gamma$-ray luminosity in the keV-MeV range. We calculate magnetic 
field in the shocks as~\citep[see, e.g.,][]{Razzaque+2004}
\begin{eqnarray}
B^\prime &=& \Big(\frac{2 \epsilon_B L_{\gamma,\rm iso}/\epsilon_e}{R_{\rm in}^2 c \Gamma^2}\Big)^{1/2} \nonumber \\
&\approx& 130~\Big(\frac{\epsilon_{B}}{0.1}\Big)^{1/2}\Big(\frac{\epsilon_{e}}{0.01}\Big)^{-1/2} \Big(\frac{L_{\gamma,\rm iso}}{10^{52}~\rm erg/s}\Big)^{1/2} \nonumber \\
&&\times \Big(\frac{R_{\rm in}}{2\times 10^{13}~\rm cm}\Big)^{-1} \Big(\frac{\Gamma}{10^3}\Big)^{-1}~{\rm kG},
\label{bfield}
\end{eqnarray}
where $\epsilon_e=0.01$ is a fraction of the shock energy that is carried by the relativistic electrons, which promptly radiate most of this energy in $\gamma$ rays (so-called fast-cooling scenario), and $\epsilon_B=0.1$ is a fraction of the shock energy that is carried by the magnetic field. 
 Magnetic energy density, $B^{\prime 2}/8\pi$, in this scenario far exceeds the radiation energy density and synchrotron radiation is
 the most effective energy loss channel for the relativistic electrons. Only a modest electron Lorentz factor
\begin{eqnarray}
\gamma_e^\prime \approx 244 \left( \frac{E_\gamma}{100~{\rm keV}} \right)^{1/2}
\left( \frac{B^\prime}{130~{\rm kG}} \right)^{-1/2} 
\left( \frac{\Gamma}{1000} \right)^{-1/2},
\label{synpk}
\end{eqnarray}
is required to explain the observed $E_\gamma \approx 100$~keV peak photon energy from GRB~130427A in this scenario.

A large fraction of the jet energy, however, is carried by the heavy nuclei in our scenario. Heavy nuclei, with atomic number $Z$, can be accelerated quickly in this magnetic field within a time scale~\citep{wang}
\begin{eqnarray}
t^\prime_{\rm acc} &=& \eta \frac{2 \pi E^\prime_A}{ZeB^\prime c} \nonumber \\
&\approx& 2\times 10^{-6} \Big(\frac{\eta}{10}\Big)\Big(\frac{Z}{26}\Big)^{-1} \Big(\frac{B^\prime}{130\rm~kG}\Big)^{-1}\Big(\frac{E_A^\prime}{\rm TeV}\Big)~{\rm s},
\label{tacc}
\end{eqnarray}
where $E_A^\prime$ is energy of the nuclei and $\eta = 10$ is a fiducial value for the number of gyroradius required for acceleration. The maximum energy is limited from comparing this time scale with the shorter of the jet dynamic time scale, given by
\begin{equation}
 t^\prime_{\rm dyn}=\frac{R_{\rm in}}{\Gamma \rm c}=0.7~\Big(\frac{R_{\rm in}}{2\times 10^{13}~\rm cm}\Big)\Big(\frac{\Gamma}{10^3}\Big)^{-1}~{\rm s},
\label{tdyn}
\end{equation}
and the energy loss time scales, which we discuss next.

Photodisintegration by interacting with the observed keV-MeV photons in the GRB jet is the main and most interesting process for energy losses by heavy nuclei. The observed differential photon flux at the Earth, $f_{\gamma}(\epsilon)$, e.g., in $\rm (MeV^{-1} s^{-1} cm^{-2})$ units, can be converted to photon density per unit energy and volume $n_\gamma^\prime(\epsilon^\prime)$, e.g., in $(\rm MeV^{-1} cm^{-3})$ units, in the comoving jet frame by the relation~\citep[see, e.g.,][]{razzaque_1},
\begin{equation}
{
n_\gamma^\prime(\epsilon^\prime) =\frac{2 d_L^2}{R_{\rm in}^2 c}
f_\gamma({\epsilon}) = \frac{2 d_L^2}{R_{\rm in}^2 c}
 f_\gamma \left( \frac{\Gamma \epsilon^\prime}{1+z} \right),
}
\label{ph_convert}
\end{equation}
where $d_L$ is the luminosity distance to the source. For GRB~130427A redshift of $z=0.34$, $d_L = 1.8$~Gpc using the standard $\Lambda$CDM cosmology.  These photons interact with the heavy nuclei by the Giant Dipole Resonance (GDR) process~\citep{pdisstek_69,pdpuget_76,pdstecker_mon99,anchor_cygob2} and the rate of such interactions is given by~\citep{stecker68},
\begin{equation} 
{t^\prime_A}^{-1} ({\gamma^\prime_A})=\frac{c} {2 {\gamma^\prime_A}^2} 
\int_{\varepsilon''_{\rm th}}^{\infty} \varepsilon''_\gamma 
\sigma_{A}(\varepsilon''_\gamma)d\varepsilon''_\gamma
\int_{\varepsilon''_\gamma/2\gamma_A^\prime}^{\infty}
\frac{n_\gamma^\prime(\epsilon^\prime)} {\epsilon^{\prime 2}} 
d{\epsilon^\prime}.
\label{tdis}
\end{equation}
Here $\gamma_A^\prime = E_A^\prime/m_A c^2$ is the Lorentz boost factor of the energetic nuclei, $\varepsilon_\gamma^{\prime\prime}=\gamma_A^\prime \epsilon^\prime(1-\beta_A\cos\theta)$ is the photon energy in the rest frame of the nuclei with an angle $\theta$ between their velocity vectors, $\varepsilon^{\prime\prime}_{\rm th}$ is the threshold photon energy for the nuclei excitation, $\beta_A \approx 1$ and $\sigma_{A}(\varepsilon^{\prime\prime}_\gamma)$ is the photodisintegration cross section. We use the threshold energy $\varepsilon^{\prime\prime}_{\rm th} = 10$~MeV and cross sections given in~\citet{puget76,anchor_cygob2,wang}. We use the same rate formula in equation~(\ref{tdis}) for the related photopion interactions with $\varepsilon^{\prime\prime}_{\rm th} = 145$~MeV per nucleon and the delta resonance cross section given in~\citet{mucke00} scaled by a factor $A^{2/3}$ for a nuclei with mass number $A$~\citep{wang}.

\begin{table}
\begin{center}
\caption{\label{table} 
Observed prompt $\gamma$-ray emission properties for GRB~130427A and corresponding Fe-photodisintegration properties. See main text for details.}
\begin{tabular}{llll}
\hline
Parameters & & Time intervals &  \\
\hline
& `$a$' & `$b$' & `$c$'  \\
\hline
Interval (s) & $- 0.1$ to 4.5 & 4.5 to 11.5 & 11.5 to 33.0 \\
Spectral model & SBPL & PL & SBPL+PL \\
$L_{\gamma,\rm iso}$ (erg/s) & $1.1\times 10^{52}$ & -- & $3.7\times 10^{51}$ (SBPL) \\
                     & -- & -- & $5.3\times 10^{50}$ (PL) \\
$E_{\gamma}$ (keV) & $10-10^5$ & $(0.3-8)\times 10^5$ & $8-5\times 10^7$ \\
                   & -- &  -- & $200-7.3\times 10^7$ \\
$\gamma_{\rm Fe}^\prime$ & $3\times 10^5-30$ & $100-4$ & $4\times 10^5-1$ \\
$\gamma_{\rm O}^\prime$ & $7\times 10^5-75$ & $250-9$ & $9\times 10^5-1$ \\
$E_{\gamma}$ (GeV) [Fe] & $9\times 10^5-90$  & $300-11$ & $10^6-3$ \\
$E_{\gamma}$ (GeV) [O] & $6\times 10^6-560$  & $2\times 10^3-70$ & 
$7\times 10^6-7$ \\
\hline
\end{tabular}
\end{center}
\end{table}

The observed prompt flux from GRB~130427A has been measured by the {\it Fermi}-GBM and {\it Fermi}-LAT in three time intervals, labelled `$a$', `$b$' and `$c$'~\citep{sci_grb_130427a} as also listed in Table~1. The $\gamma$-ray spectra in these intervals are fitted by a smoothly-broken power-law (SBPL) model (interval `$a$'), a power-law (PL) model (interval `$b$'), and a SBPL+PL model (interval `$c$') in the energy ranges as indicated by $E_\gamma$ in the table\footnote{with flux $\nu F_\nu \ge 2\times 10^{-8}$~erg~cm$^{-2}$~s$^{-1}$} (in interval `$c$' the upper line is for SBPL and the lower line is for PL). The corresponding luminosities in the intervals `$a$' and `$b$' are also listed. Note that the GBM detectors were saturated in the interval `$b$'and a full spectrum at lower energies is not available~\citep{sci_grb_130427a}. The PL component in the interval `$c$' is particularly hard, with photon index $-1.66\pm 0.13$, extending up to $\sim 73$~GeV~\citep{sci_grb_130427a}. The Table~1 also shows the 
ranges of Fe-nuclei Lorentz boost factor $\gamma^\prime_{\rm Fe}$ that will be interacting with the observed photons and the photodisintegrated $\gamma$ rays with energy 
\begin{equation}
E_{\gamma} = 2{\bar E}^{\prime\prime}_{\gamma,A} \gamma_{A}^\prime \Gamma/(1+z),
\label{Edis}
\end{equation}
where ${\bar E}^{\prime\prime}_{\gamma,\rm Fe} \sim 2$-4~MeV and ${\bar E}^{\prime\prime}_{\gamma,\rm O} \sim 5$-7~MeV~\citep{anchor_cygob2}. Thus heavy nuclei in the interval `$c$' can be disintegrated most effectively to produce a broad $\gamma$-ray spectrum in the {\it Fermi}-LAT energy range. We have used ${\bar E}^{\prime\prime}_{\gamma,\rm Fe} = 2$~MeV and ${\bar E}^{\prime\prime}_{\gamma,\rm O} = 5$~MeV in Table~1. The absence of an additional PL in the intervals `$a$' and `$b$' is compatible with inefficiency of photodisintegration in those time intervals to produce GeV $\gamma$ rays. 

\begin{figure}
\includegraphics[width=8.5 cm,height=12cm,keepaspectratio]{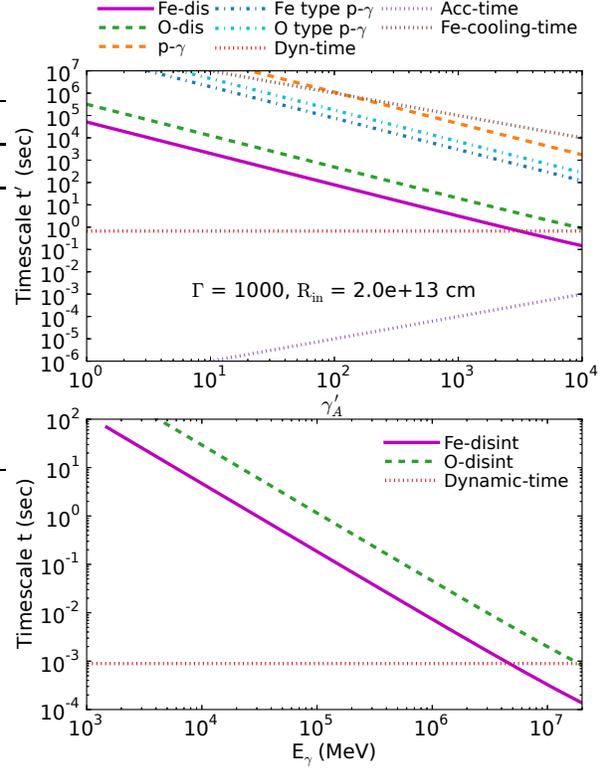}
\caption{Top: Comparison of different timescales per nucleon energy in the GRB jet frame using the properties in interval `$c$' (see Table~1). Bottom: Comparison of the observed timescales for photodisintegrated $\gamma$-ray emission for observed $\gamma$-ray energy and the dynamic time. We have used ${\bar E}^{\prime\prime}_{\gamma,A} = 1$~MeV in these plots.}
\label{fig:comp_c_Fe}
\end{figure}

The top panel of Fig.~\ref{fig:comp_c_Fe} shows the jet-frame photodisintegration timescale $t^\prime_A$ for Fe and O nuclei as functions of the Lorentz factor of the nuclei
$\gamma^\prime_A$ for the time interval `$c$'.  Also shown are the timescales for photopion ($p$-$\gamma$) interactions for Fe and O nuclei, as well as the dynamic 
time $t^\prime_{\rm dyn}$.  The acceleration timescale $t^\prime_{\rm acc}$ is shown as the dotted line with positive slope. 
Iron nuclei can be accelerated to an energy $E_{\rm Fe}\sim 3\times 10^{20}$~eV within the dynamic time scale. However, photodisintegration limits the maximum nuclei energy, from the condition $t^\prime_{A} = t^\prime_{\rm dyn}$, to  $E_{\rm Fe, max} \approx 120$~PeV and $E_{\rm O, max} \approx 110$~PeV, respectively for Iron and Oxygen. The bottom panel of Fig.~\ref{fig:comp_c_Fe} shows the photodisintegration time scale as a function of the observed secondary $\gamma$-ray energy in equation~(\ref{Edis}) with ${\bar E}^{\prime\prime}_{
\gamma,A} = 1$~MeV. Note that the Fe photodisintegration time scale in the GeV energy range matches with the time interval `$c$' in Table~1 when {\it Fermi}-LAT detected the hard power-law component in the $\sim 1$-70~GeV range. Photopion ($p$-$\gamma$) losses are not significant for our model parameters, as also found by~\citet{review_supp2} for protons in realistic GRB environment.

Apart from photodisintegration and photopion losses, heavy nuclei can also lose energy through synchrotron radiation. The time scale for this process is given by~\citep{wang}

{
\begin{eqnarray}
t^\prime_{\rm syn} &=& \frac{6 \pi m_p^4 c^3}{\sigma_T m_e^2 E^\prime_A {B^\prime}^2} \Big(\frac{A}{Z}\Big)^4 \nonumber \\
&\approx & 5.7 \times 10^{6}
\Big(\frac{B^\prime}{130~\rm kG}\Big)^{-2} 
\Big(\frac{E_A^\prime}{\rm TeV}\Big)^{-1} 
\nonumber \\ && \times 
\Big(\frac{A}{56}\Big)^4 \Big(\frac{Z}{26}\Big)^{-4}~{\rm s}.
\end{eqnarray}
This time scale for Fe nuclei is shown in Fig.~\ref{fig:comp_c_Fe} top panel labelled as ``Fe-cooling-time'' and is too long to be significant in the energy range of our interest.
}

\section{GeV $\gamma$-ray flux at the Earth}
\label{sec:gamma_flux}

The nuclear $\gamma$-ray emission (number per unit volume per unit time per unit energy)  from photodisintegration of a given nucleus can be written as~\citep[see][and references therein]{anchor_cygob2}
\begin{equation}
q_{\gamma}^\prime(E_{\gamma}^\prime) = \frac{\bar n_A m_Nc^2}{2 \bar E^{\prime\prime}_{\gamma,A}} \int_\frac{m_Nc^2 E_\gamma^\prime}{2 \bar E^{\prime\prime}_{\gamma,A}} \frac{dn_A^\prime}{dE_N^\prime} R_A^\prime(E_N^\prime) \frac{dE_N^\prime}{E_N^\prime},
\label{qg}
\end{equation}
where $\bar E^{\prime\prime}_{\gamma,A}$ is the average $\gamma$-ray energy from the nuclei de-excitation when it is assumed that the $\gamma$-ray spectrum is monochromatic, and $\bar n_A$ is the average multiplicity of these $\gamma$ rays, with ${\bar n}_{\rm Fe} = 1$-3 and ${\bar n}_{\rm O} = 0.3$-0.5~\citep{anchor_cygob2}. The first term inside the integration in equation~(\ref{qg}), $dn^\prime_{A} / dE_N^\prime = A_{N}^\prime E_N^{\prime -\alpha}$ is the nucleon spectrum per nucleon energy, with $A^\prime_N$ being a normalization constant. Of course $dn_{A}^\prime /dE_A^\prime = (1/A) dn_{A}^\prime /dE_N^\prime$ in terms of the nucleus energy $E_A^\prime$. The second term inside the integration in equation~(\ref{qg}), $R_A^\prime(E_N^\prime) = t^{\prime -1}_{A}$, with $E_N^\prime = \gamma^\prime_A m_Nc^2$, is the scattering rate of the GDR interactions.

The emission coefficient (number per unit volume per unit time) of the photodisintegration $\gamma$-rays in the wind rest frame is $j_{\gamma}^\prime(E_{\gamma}^\prime) = E_\gamma^\prime q_{\gamma}^\prime(E_{\gamma}^\prime)$, which can be related to the emission coefficient in the lab frame as $j_{\gamma}^*(E_{\gamma}^*)  = (E_\gamma^* / E_\gamma^\prime)^2 j_{\gamma}^\prime (E_{\gamma}^\prime) = \Gamma^2  j_{\gamma}^\prime (E_{\gamma}^*/\Gamma)$~\citep{Rybicki_Lightman}. Similarly the four volume invariance gives the volume in the lab frame as $ V^* = \Gamma V^\prime$~\citep{Dermer_Menon2009}, where $V^* = 4\pi R_{\rm in}^3/3$. The flux in the lab frame is therefore given by
\begin{equation}
E_\gamma^* f_{\gamma, A} (E_\gamma^*) = \frac{j^* (E_\gamma^*) V^*}{4\pi d_L^2}
= \frac{\Gamma E_\gamma^* V^*}{4\pi d_L^2} q_\gamma^\prime (E_\gamma^*/\Gamma),
\end{equation}
and finally the differential flux in, e.g., (MeV$^{-1}$~s$^{-1}$~cm$^{-2}$) units at the Earth is given by
\begin{equation}
f_{\gamma, A} (E_\gamma) = \frac{\Gamma^2 V^\prime}{4\pi d_L^2} 
q_\gamma^\prime \left( \frac{E_\gamma (1+z)} {\Gamma} \right).
\label{disflux}
\end{equation}

\begin{figure}
\includegraphics[width=8.5 cm,height=12cm,keepaspectratio]{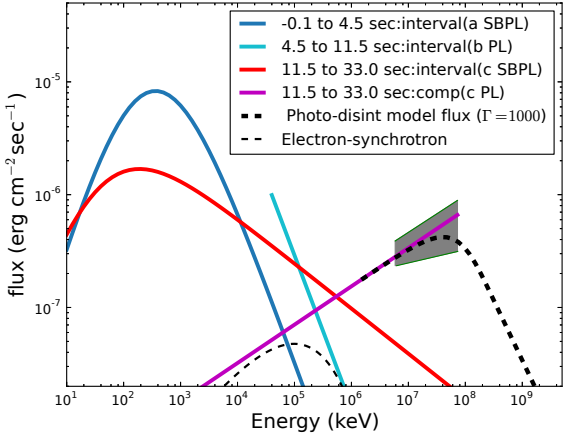}
\caption{Observed prompt $\gamma$-ray energy flux (solid lines) from GRB~130427A in 3 different time intervals~\citep{sci_grb_130427a}. The thick-dashed line corresponds to our fit to the PL component (photon index $-1.66\pm 0.13$) in the interval `$c$' using the Fe-photodisintegration model in equation~(\ref{disflux}). The turnover of the model flux above $\sim 70$~GeV is due to the $\gamma\gamma \to e^\pm$ pair production with the low-energy photons in the SBPL component. The uncertainty in the observed PL flux component, where it dominates over the SBPL component, is shown with shaded region.
 The thin-dashed line corresponds to the synchrotron flux from the secondary $e^\pm$ pairs.}
\label{fig: obsflux}
\end{figure}

In Fig.~\ref{fig: obsflux} we fit the photodisintegration model flux (thick-dashed line) in equation~(\ref{disflux}) for the Iron nuclei, with parameters ${\bar n}_{\rm Fe} = 2$ and $\bar E^{\prime\prime}_{\gamma,\rm Fe} = 1$~MeV, to the observed additional PL component in the interval `$c$' during the prompt emission phase of GRB~130427A~\citep{sci_grb_130427a}. To model the $\sim 1$-70~GeV component in the interval `$c$' we need a parent Fe-nuclei spectrum in the GRB jet-frame which is given by
\begin{equation}
\frac{dn_{\rm Fe}^\prime}{dE_{\rm Fe}^\prime}= 1.3 \times 10^{20} {E_{\rm Fe}^\prime}^{-3.05}
~\rm cm^{-3}~MeV^{-1} .
\label{Feflux}
\end{equation}
This corresponds to an isotropic-equivalent luminosity of the Fe-nuclei in the jet of GRB~130427A as given by
\begin{eqnarray}
L_{\rm Fe, iso} &=& 4\pi R_{\rm in}^2 c\Gamma^2 
\int^{E^\prime_{\rm Fe, max}}_{\frac{E^\prime_{\gamma} m_{\rm Fe}c^2} {2 {\bar E}^{\prime\prime}_{\gamma,\rm Fe}}}  E^\prime_{\rm Fe} 
\frac{dn_{\rm Fe}^\prime}{dE_{\rm Fe}^\prime}
dE^\prime_{\rm Fe} \nonumber \\
&\approx & 3.2\times 10^{53} ~{\rm erg}~{\rm s}^{-1},
\end{eqnarray}
where the lower limit of the integration follows from the equation~(\ref{Edis}). As for comparisons, the isotropic-equivalent luminosity in the PL component of 1-70~GeV $\gamma$-rays is $\approx 5\times 10^{50}$~erg/s, i.e., about three orders of magnitude lower. The isotropic-equivalent $\gamma$-ray luminosity in the SBPL component in the same time-interval `$c$' is $\approx 4 \times 10^{51}$~erg/s, which is less than two orders of magnitude lower than the luminosity in the Fe nuclei. However, $L_{\rm Fe, iso}$ is of the same order as the peak $\gamma$-ray luminosity of $L_{\gamma,\rm iso} = 2.7\times 10^{53}$~erg/s~\citep{Maselli+2014} 
 and the magnetic luminosity, $L_B = 4\pi R_{\rm in}^2 c (B^{\prime 2}/8\pi)\Gamma^2 \approx 10^{53}$~erg/s. The total energy in the relativistic Fe nuclei is $E_{\rm Fe, iso} \approx \Delta t_cL_{\rm Fe, iso}/(1+z)\approx 5\times 10^{54}$~erg, where $\Delta t_{c} = 21.5$~s is the interval `$c$' duration. A comparison with the total isotropic-equivalent $\
gamma$-ray energy released from the GRB~130427A, $E_{\gamma,\rm iso} = 8.1\times 10^{53}$~erg~\citep{Maselli+2014}, we find that $E_{\rm Fe, iso}$ is only a factor $\lesssim 10$ times higher. This is a very conducive situation for the hadronic model from the total energy perspective.

The maximum $\gamma$-ray energy from Fe-disintegration is $E_\gamma \approx 5$~TeV for $E_{\rm Fe, max} \approx 120$~PeV, following equation~(\ref{Edis}) with $\bar E^{\prime\prime}_{\gamma,\rm Fe} = 1$~MeV. Photons of this energy cannot escape the GRB jet due to $e^\pm$ pair creation by interacting with the same low-energy photons in the SBPL component which are responsible for photodisintegration. We calculate the $\gamma\gamma\to e^\pm$ pair-production opacity, following \citet{Gould+1967}, as
\begin{eqnarray}
\tau_{\gamma\gamma} (E_\gamma) &=& \frac{R_{\rm in}}{\Gamma} 
\pi r_0^2 \left[ \frac{m_e^2 c^4 \Gamma}{(1+z)E_\gamma} \right]^2 \nonumber \\
&& \times \int_{\frac{m_e^2c^4\Gamma}{(1+z)E_\gamma}}^{\frac{(1+z)E_\gamma}{\Gamma}}
\frac{n^\prime (\epsilon^\prime)}{\epsilon^{\prime 2}}
\varphi [S_0 (\epsilon^\prime)] d\epsilon^\prime,
\label{ggopt}
\end{eqnarray}
where $r_0$ is the classical electron radius and $n^\prime (\epsilon^\prime)$ is the isotropic distribution of photons in the GRB jet given by the equation~(\ref{ph_convert}). The function $\varphi [S_0 (\epsilon^\prime)]$, with $S_0 (\epsilon^\prime) = (1+z)\epsilon^\prime E_\gamma/\Gamma m_e^2c^4$, is defined by \citet{Gould+1967} and corrected by \citet{Brown+1973}. For the fiducial values of $\Gamma = 10^3$ and $R_{\rm in} = 2\times 10^{13}$~cm, we calculate $\tau_{\gamma\gamma} \approx 1$ at $E_\gamma  \approx 70$~GeV for the time-interval `$c$' in Table~1 for GRB~130427A. Figure~\ref{fig: obsflux} shows the photodisintegrated $\gamma$-ray flux model (thick-dashed curve) using the opacity in equation~(\ref{ggopt}) to modify the flux in equation~(\ref{disflux}) in the slab approximation~\citep{Dermer_Menon2009}.  

The secondary $e^\pm$ pairs from $\gamma\gamma$ interactions will be produced with a maximum energy of $E^\prime_e \approx {\bar E}^{\prime\prime}_{\gamma,\rm Fe}\gamma_{\rm Fe}^\prime \approx 3$~GeV. The characteristic synchrotron photon energy from these pairs, with $B_Q = 4.414\times 10^{13}$~G, is given by
\begin{eqnarray}
E_{\gamma,\rm syn} &=& 
\frac{3B^\prime}{2B_Q}\gamma_e^{\prime 2} m_ec^2 \frac{\Gamma}{1+z}
\nonumber \\
&\approx& 60 \left( \frac{B^\prime}{130~{\rm kG}} \right) 
\left(\frac{\Gamma}{10^3}\right)~{\rm MeV}.
\label{synenergy}
\end{eqnarray}
 The $e^\pm$ pairs are essentially in the fast-cooling regime, and their synchrotron emission flux is shown (thin-dashed line) in Fig.~\ref{fig: obsflux}, which
 is below the primary SBPL component.

\section{Beta-decay neutrinos}
\label{sec:neutrino}

A secondary neutron is produced approximately every other GDR interactions, see equation~(\ref{photo}), which will decay to produce an electron antineutrino. Due to relativistic effect~\citep{Razzaque+2006}, the time scale for the neutron decay neutrino emission in the observer frame is of the same order as the interval `$c$'. Since the multiplicity of photodisintegrated $\gamma$ rays is 2 and the energy of ${\bar \nu}_e$ is roughly 1/2 of the energy of the $\gamma$-ray in equation~(\ref{Edis}) as given by~\citet{anchor_cygob2}, the neutrino source flux can be estimated as $f_{{\bar \nu}_e} (E_\nu) = (1/2)f_{\gamma,A} (2E_\gamma)$. 

After neutrino flavor oscillation over cosmological distance, the flux will be modified and the oscillation probability can be written for a given production flavor $\alpha$ and an observable flavor $\beta$ as $P (\nu_\alpha \to \nu_\beta) = \sum_{i} \left| U_{\beta i}\right|^2 \left| U_{\alpha i}\right|^2$, where $U_{\alpha i}$ is the PMNS mixing matrix~\citep{PDG2014}. The fluxes on the Earth are modified by this probability as $f_{\beta}  (E_\nu) = P (\nu_\alpha \to \nu_\beta) f_{\alpha} (E_\nu)$. For the current best-fit oscillation parameter values~\citep{Fogli+2012} the ratios of fluxes of different flavors at the source,  $f_{{\bar \nu}_e} : f_{{\bar \nu}_\mu} : f_{{\bar \nu}_\tau} = 1 : 0 : 0$ from the beta decay neutrinos, will be modified to $f_{{\bar \nu}_e} : f_{{\bar \nu}_\mu} : f_{{\bar \nu}_\tau} \approx 0.55 : 0.27 : 0.18$, at the Earth.

We calculate the number of neutrino events expected from GRB~130427A in our photodisintegration model, using the IceCube effective area in~\citet{geV_ne_csection} scaled to the full IceCube, as
\begin{equation}
N_{\bar \nu_\mu} =\frac{f_{{\bar \nu}_\mu}}{f_{{\bar \nu}_e}} 
\frac{\Delta t_{c}}{2}\int_{100~\rm GeV}^{2.5~\rm TeV}
 A_{\rm eff}(E_{\nu}) f_{\gamma,\rm Fe}(2E_{\nu}) dE_{\nu}.
\end{equation} 
The expected neutrino events from GRB~130427A in our model is $N_{\bar \nu_\mu}\approx 0.003$, which is consistent with IceCube non-detection~\citep{abbasi_no_grbnu}. Therefore no constraints can be derived on our model, as was done by~\citet{noneutrinogao} for a primary $p$-$\gamma$ model, using non-detection of neutrinos from GRB~130427A.

\section{Summary and conclusions}
\label{sec:conclusion}

GRB~130427A is one of the brightest and energetic GRBs detected at a relatively low redshift of 0.34. Detection of a hard PL component in the prompt phase by the {\it Fermi}-LAT, that extends up to 73~GeV, is very interesting and beg explanation of its origin. We have modeled this spectral component using photodisintegrated $\gamma$-ray emission from an Iron-rich GRB jet. The observed keV-MeV $\gamma$ rays, presumably produced by synchrotron radiation from primary electrons or by another process, provide the necessary target photons required for this model. The time scale required for Fe photodisintegration with target photons is compatible with the 11.5-33.0~s interval of GRB~130427A when the hard PL spectral component was detected by the {\it Fermi}-LAT. Non-detection of a PL component at earlier times can be interpreted as inefficiency of the photodisintegration process to produce $\gamma$ rays in the LAT range.

The total isotropic-equivalent energy required in the relativistic Fe nuclei in this model, $\approx 5\times 10^{54}$~erg, is relatively modest for a hadronic model and is only $\lesssim 10$ times the total isotropic-equivalent energy released in $\gamma$ rays from GRB~130427A. This is due to relatively higher efficiency of the photodisintegration process than other hadronic processes, such as the photopion production. A more challenging issue, however, is to explain the origin of heavy nuclei in the GRB jet. 
\citet{wang} suggested a possibility that the GRB jet could be rich in heavy nuclei initially, as the subrelativistic jet deep inside the GRB progenitor star can 
entrap core material~\citep{Zhang+2003}. This is the scenario we have adopted to explain the observed prompt GeV $\gamma$-rays by Fe disintegration.
In case the Fe is a subdominant component in the GRB jet, the required total energy will increase. For example, in case of solar composition the jet will 
be dominated by proton 
and light nuclei and the total jet energy will need to be $\sim 10^3$ times the value we calculate, if all GeV $\gamma$ rays originate from Fe disintegration only. 
However, photodisintegration of intermediate and light nuclei will also contribute to the observed $\gamma$-ray flux in such a case and the total energy requirement
may not be so severe. A detailed study will be presented elsewhere. We believe a phenomenological approach such as ours to explain observations with the model 
proposed can give clues to the yet unknown origin of the prompt GeV emission and ultimately clues to the composition of the GRB jets.

\section*{Ackowledgments}

This work was supported in part by the National Research Foundation (South Africa) grants no.\ 87823 (CPRR) and no.\ 93273 (MWGR) to SR. JCJ is thankful to SR and the Raman Research Institute for supporting a visit to the University of Johannesburg where most of this work was done.

\footnotesize{
\bibliography{ref}
}

\end{document}